\documentclass[11pt]{article}

\usepackage{asp2014}

\aspSuppressVolSlug
\resetcounters

\begin{document}

\title{Prospects about X- and gamma-ray counterparts of gravitational wave signals with INTEGRAL}
\author{P.~Bacon, V.~Savchenko, E.~Chassande-Mottin and P.~Laurent}
\affil{APC, AstroParticule et Cosmologie, Universit\'e Paris Diderot, CNRS/IN2P3, CEA/Irfu,
Observatoire de Paris, Sorbonne Paris Cit\'e, Paris, France; \email{philippe.bacon@apc.in2p3.fr}}

%% This section is for ADS Processing.  There must be one line per author.
\paperauthor{P.~Bacon}{philippe.bacon@apc.in2p3.fr}{}{APC, Astroparticule et Cosmologie CNRS/IN2P3}{Physics}{Paris}{}{75013}{France}
\paperauthor{V.~Savchenko}{savchenk@apc.in2p3.fr}{}{APC, Astroparticule et Cosmologie CNRS/IN2P3}{Physics}{Paris}{}{75013}{France}
\paperauthor{E.~Chassande-Mottin}{ecmn@apc.in2p3.fr}{0000-0003-3768-9908}{APC, Astroparticule et Cosmologie CNRS/IN2P3}{Physics}{Paris}{}{75013}{France}
\paperauthor{P.~Laurent}{philippe.laurent@apc.in2p3.fr}{}{APC, Astroparticule et Cosmologie CNRS/IN2P3}{Physics}{Paris}{}{75013}{France}

\begin{abstract}
  By extrapolating the number of detections made during the first LIGO science
  run, tenths of gravitational wave signals from binary black hole mergers are
  anticipated in upcoming LIGO Virgo science runs. Finding an electromagnetic
  counterpart from compact binary merger events would be a capital achievement.
  We evaluate the ability of current wide-field X- and gamma-ray telescopes
  aboard INTEGRAL to find such counterparts thanks to an end-to-end simulation
  for estimating the fraction of the sources that can be followed up and/or
  detected.
\end{abstract}

On September 14, 2015, the two LIGO interferometers realized the first direct detection of
gravitational waves (GW). A next step would be to associate an electromagnetic (EM)
counterpart to a GW event. We address the question of the joint detectability 
of GW events by Advanced LIGO, and associated EM events at high-energies by the 
INTEGRAL mission. This study is similar to that of \citep{2016arXiv160606124P} done with Fermi.

\section*{Monte-Carlo simulation of gravitational-wave events from neutron-star binary mergers}

We produced a simulated catalogue of 6 millions Milky Way-like galaxies
distributed to $z\sim 0.12$. To those galaxies, we associate binary neutron star
(BNS) mergers with same mass distribution and rate $R=23.5 \mathrm{Myr}^{-3}$ per galaxy
according to Model A (solar metallicity) of
\citep{2012ApJ...759...52D}. This results in a population of 14\,000
mergers for a $100$-year simulation.

The GW signals corresponding to the binary mergers was simulated using the
\texttt{TaylorT4} model and injected into Gaussian noise coloured according to
best sensitivity curves anticipated for the 2nd LIGO/Virgo run O2
\citep{2016LRR....19....1A}. We distribute those events randomly in time over
the expected period of O2 from Nov 2016 to May 2017. We recover the signal with
a simplified matched-filtering algorithm \citep{2016PhRvD..93b4013S} in the case
of the LIGO-only two-detector network, and that of the LIGO-Virgo three-detector
network. We select the GW events with $\mathrm{SNR}_{\mathrm{combined}} > 9$
corresponding to a false-alarm rate $\mathrm{FAR} < 1/\mathrm{month}$.  With
this selection cut, BNS merger can be recovered down to an horizon distance of
about $150$ Mpc and $50$ Mpc for the LIGO and Virgo detectors respectively.
This results in 170 (resp. 192) selected BNS merger signals over 100
years which corresponds to $\sim 1$ mergers over the 6-month duration of O2.

We recover GW signals using matched-filtering techniques and reconstruct the
posterior skymaps giving the position of the source using the BAYESTAR pipeline
\citep{2016PhRvD..93b4013S}.  With two detectors, the position uncertainty is
contained in a window of 700-2100 square degrees, while with three detectors, it
is appreciably diminished to a 300-1700 square-degree window. The latter
estimate is obtained using the 70 \% fraction of events where the position
estimation converged.

\section*{Electromagnetic emission model from neutron-star binary mergers}

Compact binary mergers involving at least one neutron star are considered to be
plausible progenitors of short gamma-ray bursts (GRB). Short GRBs is a
subset of the whole GRB population with duration shorter than $\sim$2
seconds. Short GRBs have also a harder spectrum. Prompt emission is attributed
to dissipation within the collimated ultra-relativistic jet and is therefore
strongly beamed. It is followed by slowly decaying afterglow, observed from
radio to GeV, produced by the interaction between the ejecta and the ambient
medium surrounding the GRB progenitor. Afterglow emission is originally also
strongly beamed, but as the outflow decelerates, the emission becomes
progressively more isotropic.

In our simulation we assume that every BNS merger is associated with a short
gamma-ray burst. We simulate synthetic prompt GRB and a hard X-ray afterglow,
assuming the following simplified model. 

\subsection*{Prompt emission}

We model the prompt emission spectrum by the cut-off power-law model. This model
is a variant of the Band model \citep{1993ApJ...413..281B} where high-energies
have been suppressed because they are generally not well constrained by the
observations. For simplicity we assume the same prompt emission spectrum for all
synthetic GRB, with $\alpha_{\mathrm{BAND}} = - 0.5$ and $E_{\mathrm{peak}} =
600$ keV, close to the average short GRB spectrum observed by Fermi/GBM
\citep{2014ApJS..211...12G}. We also assume fixed duration of the prompt
emission of $1 \, s$ and fixed isotropic equivalent luminosity of
$10^{50}~$erg~cm$^{-2}$~s$^{-1}$.

The principal source of uncertainty on rate of GRB detections associated with
the BNS mergers is the beaming factor of the prompt emission. We assume a fixed
beaming angle of 10 degrees. This angle is roughly compatible with the observations
of the total short GRB rate inferred from our BNS population. 

\subsection*{Hard X-ray afterglows}

Although afterglows in the hard X-rays are rarely observed, it is not
unreasonable to assume that they are as frequent as gamma-rays, but observed
much less frequently due to observational limitations. The INTEGRAL satellite
(see next section) is well-suited for the observation of hard X-ray
afterglows. There are indications that such afterglows are present in the most
luminous GRBs occurring about once in few months (2\% of the sky observed by
INTEGRAL). %For instance, GRB120711A was seen by ISGRI and JEM-X for about 8
%hours \citep{2014A&A...567A..84M}. The bright hard X-ray emission afterglow
%found to be related to the bright gamma-ray afterglow, observed by Fermi/LAT in
%the energy range above 100 MeV with a consistent power-law spectrum with index
%-2.

In this study, we assume that every BNS merger produces a hard X-ray
afterglow with a spectrum similar to GRB120711A, but rescaled to the
total EM luminosity of a typical short GRB -
$10^{50}~$erg$^{-2}~$s$^{-1}$. We extrapolate this spectrum to the
off-axis case using Model~I in \citep{2016arXiv160606124P}, assuming
$\Gamma_0$=200 and $t_{dec}=100~$s.

\section*{The INTEGRAL mission and possible follow-up strategies}

\begin{figure}
  \centering
    \includegraphics[scale=.6]{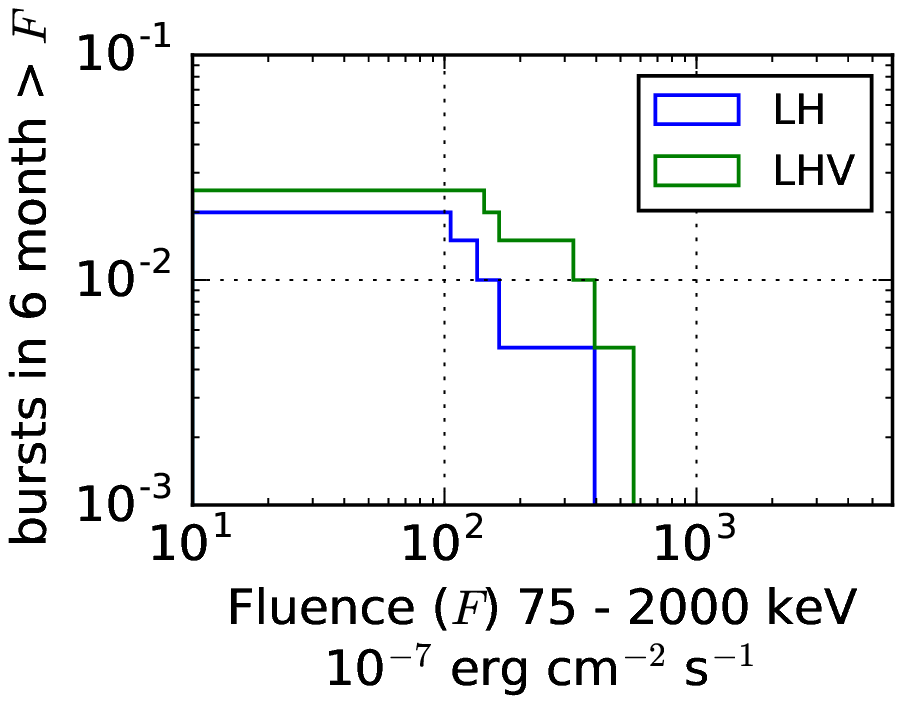}
    \includegraphics[scale=.6]{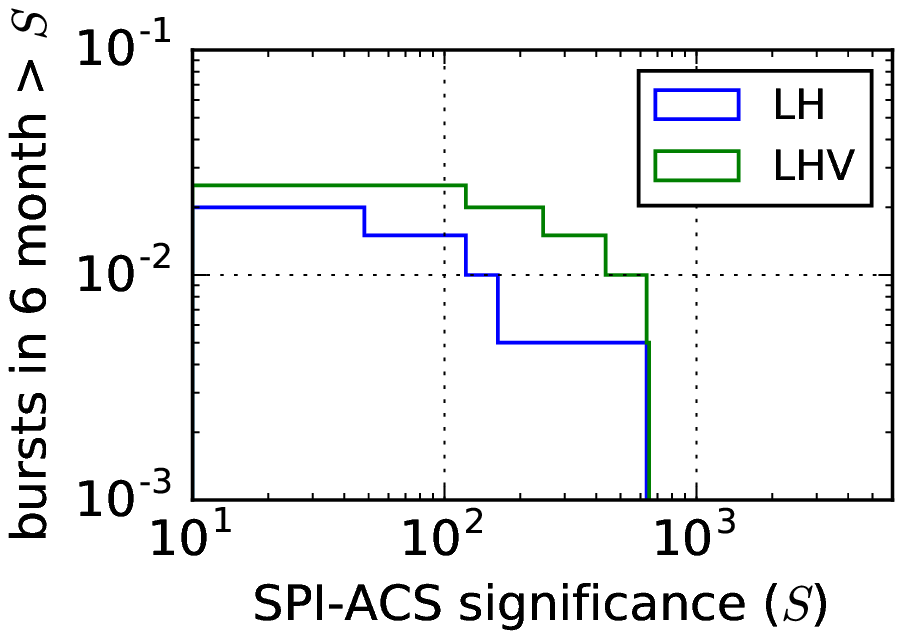}\\
    \includegraphics[scale=.6]{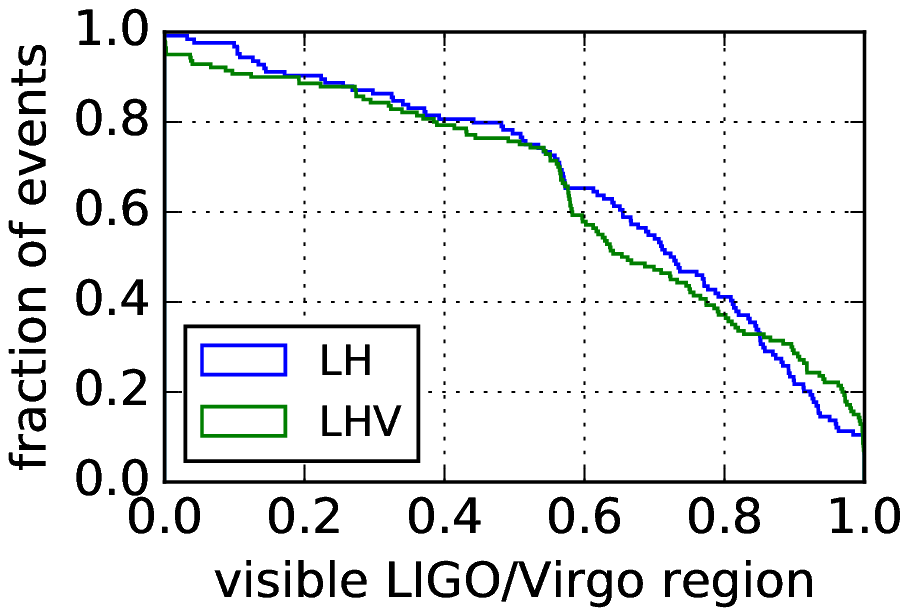}
    \includegraphics[scale=.6]{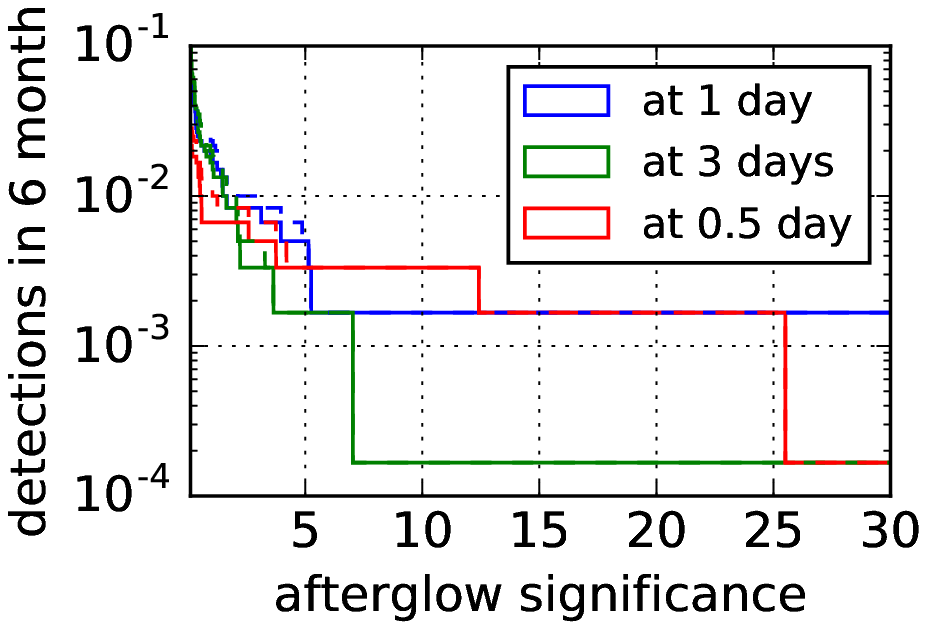}
    \caption{(\textit{Top--left}) Cumulative distribution of the detected
      fluences by INTEGRAL in the LIGO/Virgo O2 run.(\textit{Top--right})
      Cumulative distribution of the significance in the INTEGRAL/SPI-ACS
      instrument.  (\textit{Bottom--left}) Cumulative distribution of the
      fraction of the LIGO/Virgo localization region accessible to INTEGRAL
      pointed follow-up observation in the LIGO/Virgo O2
      run. (\textit{Bottom--right}) Significance distribution on detected
      sources depending on time delay after the prompt emission, only in the case of joint LIGO/Virgo operation.}
    \label{fluence_significance}
\end{figure}

The INTEGRAL satellite \citep{2003A&A...411L...1W} carries a combination of pointing
instruments (IBIS, JEM-X, OMC and SPI), covering wide
energy range from 3 keV to 10 MeV. Low-energy subsystem of IBIS, - ISGRI can achieve a sensitivity of $3.7 \times 10^{-11}$~erg~cm$^2$~s in 20-60 keV in an area of about 1000~deg$^{2}$,
with a continuous 100~ks exposure, reaching deeper than any other hard X-ray
telescope in such an extended sky region. SPI and IBIS are associated with
SPI-ACS and IBIS/Veto that are shields made of BGO scintillator with
sufficiently collecting area that they can be considered as instruments in their
own right.  This allows a \textit{passive} follow-up using the near all-sky
instruments SPI-ACS and IBIS-Veto and an \textit{active} follow-up where the
satellite is repointed.  In this study, we assume that every GW alerts is
followed actively.  Typical latency of INTEGRAL observations is about 1 day but
may be lower in some circumstances. We compute the flux at several moments of
time, and compare it with ISGRI sensitivity in 100~ks exposure. \\

%\subsection*{Prompt -- Passive follow-up}

Choosing an isotropic sky distribution for the passive follow-up
emission (prompt emission) of our GW events, we show that in every
case where the Earth is within the GRB emission cone, INTEGRAL detects
prompt emission with significance at least 40~sigma. In the LIGO LH
configuration (or LIGO/Virgo LVH) the rate of detection is 0.02
(0.025) in the 6 months of O2 for GW alerts ($\mathrm{SNR} \geq
9$). Using SPI-ACS, IBIS/Veto, and IBIS it is possible to derive a
crude localization of these events. The accuracy of this localization
strongly depends on the true source location, and ranges from 5\% to
about 50\% of the sky. Combining this observation with LIGO/Virgo
localization can, in some cases, result in a much constrained sky
region than the LIGO/Virgo observation alone. \\

%\subsection*{Afterglow -- Instrument repointing}

If we want to follow-up the afterglow emission, we will need to
re-point the on-board instruments. INTEGRAL can point only to a
fraction of the GW localization region. For each event we compute the
total localization probability within the observable region, and find
that in 40\% of the cases 80\% of the localization is
observed. However, if the complete available region is followed up
every time, in total 80\% of the true source locations are
covered. Adding Virgo results in a substantial increase of the number
of well-localized triggers. Although these triggers still constitute a
small fraction of the whole LIGO/Virgo detection rate, they provide a
unique opportunity to perform mutliwavelength observations of the
whole localization region of the GW source.

Using the same assumptions as in the case of the prompt emission, We find that
the expected rate of 5~sigma detections is only about 0.001 events in 6 months
of O2. However, the rate of detections is very sensitive to the GRB energetics.

We therefore focus on another possibility, that 10\% of the events launch much
more energetic outflow, in agreement with the luminosities observed in real
short GRBs. In this case we expect about 0.01 events seen in 6 months. We
stress that the afterglow prediction is particularly sensitive to assumptions on
rather uncertain source model, and we assume only one, yet realistic, option.

% The temporal profile of the afterglow greatly depends on the angle between the
% jet and the line of sight: light curves of the source observed at larger angles
% reach their maximum at a later moment. Assuming the jet is aligned with the
% binary rotation axis, the measurement of the inclination from the LIGO/Virgo
% data can be used to guide the follow-up observation. However, at these early
% stages of LIGO and Virgo operations binary inclination cannot be measured with
% sufficient precision.

Afterglows can be seen from binaries at a wider range of inclination
angles than the prompt emission. While we cannot reliably characterize
the shape of the observed binary inclination distribution due to
limited statistics, in our sample, equivalent to 200 times the
projected LIGO O2 run, we identify 5 synthetic hard X-ray afterglow
counterparts, among them 2 with the inclination angle larger than 10
degrees which would represent so-called ``orphan afterglows''.

\vspace{3mm}
\noindent {\normalsize \textbf{Acknowledgements}}

%----------------------------------------------------------------------------------------
%	ACKNOWLEDGEMENTS
%----------------------------------------------------------------------------------------

{\linespread{0.1} \footnotesize \noindent This research was supported
  European Union's Horizon 2020 research and innovation programme
  under grant agreement No 653477. The authors would like to thank the
  LIGO Scientific Collaboration and Virgo Collaboration for giving
  access to the simulation software used here, Barbara Patricelli for
  sharing her data with us. We are grateful to the Fran\c{c}ois Arago
  Centre at APC for providing computing resources, and to VirtualData
  from LABEX P2IO for enabling access to the academic cloud infrastructure.}

%----------------------------------------------------------------------------------------
%	REFERENCES
%----------------------------------------------------------------------------------------

\vspace*{-5mm}
{\footnotesize
\bibliographystyle{asp2014.bst} % Plain referencing style
\bibliography{P7-1.bib} % Use the example bibliography file sample.bib

\begin{thebibliography}{}
\expandafter\ifx\csname natexlab\endcsname\relax\def\natexlab#1{#1}\fi
\expandafter\ifx\csname url\endcsname\relax
  \def\url#1{\texttt{#1}}\fi
\expandafter\ifx\csname urlprefix\endcsname\relax\def\urlprefix{URL }\fi
\providecommand{\eprint}[2][]{\url{#2}}

\bibitem[{{Abbott} et~al.(2016)}]{2016LRR....19....1A}
{Abbott}, B.~P., et~al. 2016, Living Reviews in Relativity, 19.
  \eprint{1304.0670}

\bibitem[{{Band} et~al.(1993)}]{1993ApJ...413..281B}
{Band}, D., et~al. 1993, \apj, 413, 281

\bibitem[{{Dominik} et~al.(2012)}]{2012ApJ...759...52D}
{Dominik}, M., et~al. 2012, \apj, 759, 52. \eprint{1202.4901}

\bibitem[{{Gruber} et~al.(2014)}]{2014ApJS..211...12G}
{Gruber}, D., et~al. 2014, \apjs, 211, 12. \eprint{1401.5069}

\bibitem[{{Patricelli} et~al.(2016)}]{2016arXiv160606124P}
{Patricelli}, B., et~al. 2016, ArXiv e-prints. \eprint{1606.06124}

\bibitem[{{Singer} \& {Price}(2016)}]{2016PhRvD..93b4013S}
{Singer}, L.~P., \& {Price}, L.~R. 2016, \prd, 93, 024013. \eprint{1508.03634}

\bibitem[{{Winkler} et~al.(2003){Winkler}, {Courvoisier}
  et~al.}]{2003A&A...411L...1W}
{Winkler}, C., {Courvoisier}, T.~J.-L., et~al. 2003, \aap, 411, L1

\end{thebibliography}
}

\end{document}